\documentclass[aps,prl,twocolumn,showpacs,superscriptaddress,floatfix,nofootinbib]{revtex4}

\usepackage{graphicx}
\usepackage{epsfig}
\usepackage[english]{babel}

\newcommand{\be}{\begin{equation}}
\newcommand{\bea}{\begin{eqnarray}}
\newcommand{\eea}{\end{eqnarray}}
\newcommand{\ee}{\end{equation}}

\newcommand{\ket}[1]{\mbox{$| #1 \rangle$}}

\newcommand{\proj}[1]{\mbox{$|#1\rangle \!\langle #1 |$}}

\def\B{{\cal B}}
\def\tr{ \mbox{tr}}

\begin{document}

\title{Fine-grained entanglement loss along renormalization group flows}

\author{J.I. Latorre} 
\affiliation{Dept. d'Estructura i Constituents de la Mat\`eria, Univ. Barcelona, 08028, Barcelona, Spain.}
\author{C.A. L\"utken} 
\affiliation{\it Dept. of Physics, Univ. of Oslo, P.O.Box 1048 Blindern, NO-0316 Oslo, Norway}
\author{E. Rico} 
\affiliation{Dept. d'Estructura i Constituents de la Mat\`eria, Univ. Barcelona, 08028, Barcelona, Spain.}  
\author{G. Vidal}
\affiliation{Institute for Quantum Information, California Institute for Technology, 
Pasadena, CA 91125 USA} 
\date{\today}

\begin{abstract}

We explore entanglement loss along renormalization group trajectories as a basic quantum information property underlying their irreversibility. This analysis is carried out for the quantum Ising chain as a transverse magnetic field is changed. We consider the ground-state entanglement between a large block of spins and the rest of the chain. Entanglement loss is seen to follow from a rigid reordering, satisfying the majorization relation, of the eigenvalues of the reduced density matrix for the spin block.  More generally, our results indicate that it may be possible to prove the irreversibility along RG trajectories from the properties of the vacuum only, without need to study the whole hamiltonian.

\end{abstract}


\maketitle

A very exciting perspective in quantum information science is that of obtaining new insights into the properties of strongly correlated quantum many-body systems and quantum field theory from recent progress in the study of multipartite quantum entanglement \cite{QI,Pres,VLRK02}. In this letter, we shall explore the possibility, suggested in \cite{Pres,VLRK02}, of relating the irreversibility of renormalization group (RG) flows to the entanglement properties of the vacuum state of a physical system. 

Succesive RG transformations
applied to the hamiltonian of a system produce
a flow in coupling space as we 
analyze longer distances or, equivalently, smaller energies \cite{WK74}. 
Every point along this RG flow  
provides the appropriate effective hamiltonian
suited to compute all observables at a given physical scale. 
The flow is irreversible for unitary, Poincar\'e invariant,
renormalizable field theories in one dimension according to the c-theorem
 \cite{Zam86,us}.
This result may naively seem obvious since
the integration of short-distance degrees
of freedom appears to drop information. Yet, limit
cycles are known to exist for exotic theories
\cite{Si04} and
the precise hypothesis sustaining the c-theorem 
are of relevance. We envisage that the study of entanglement along RG trajectories will eventually lead to an alternative proof of their irreversibility, one based on information theoretical arguments. In the meantime, our results already show that marked vestiges of irreversibility are present in properties of the vacuum alone ---that is, in properties that, in sharp contrast with the quantities used in the c-theorem, do not involve the whole hamiltonian of the system.

The key idea in the present analysis is that a loss of entanglement occurs along RG trajectories \cite{VLRK02}. This will be discussed for the ground state $\ket{0}$ of a quantum spin chain called the XY model, with hamiltonian \cite{Pfe70,BmC71,Sa99}
\be
H=\sum_{i=1}^N \left( \frac{1+\gamma}{2}\sigma^x_i\sigma^x_{i+1}+
  \frac{1-\gamma}{2}\sigma^y_i\sigma^y_{i+1}
 + \lambda \sigma^z_i \right)
\ee
in the limit of an infinite chain,  $N\rightarrow \infty$. Most of the discussion will be conducted for the quantum Ising chain, $\gamma=1$, with an arbitrary transverse magnetic field $\lambda \in [0,\infty)$. At the critical value $\lambda^*=1$, the ground state undergoes a quantum phase transition, while the departures  $\lambda>\lambda^*$ 
and $\lambda <\lambda^*$ are both 
related to relevant operators
that drive the RG flow away. This simple model 
corresponds in the continuum to a massive fermion
whose mass is monotonically related to $\vert \lambda-1\vert$. 
The reduced density matrix $\rho_L  \equiv \tr_{\neg {\B}_L} \proj{0}$ for a block 
$\B_L$ of $L$ contiguous spins, can be computed using the techniques developed in Ref.~\cite{VLRK02}. 
The von Neumann entropy of this mixed state,
\be
S_L(\gamma,\lambda) \equiv - \tr \left( \rho_{L} 
\log_2 \rho_{L} \right), 
\label{eq:SL}
\ee
quantifies how entangled the block $\B_L$ is with the rest of the spin chain. In \cite{VLRK02}, the dependence of this entropy in the size $L$ of the block was analyzed, revealing that a saturation value $S(\gamma,\lambda)$ is achieved for block sizes $L$ larger than the correlation length $\xi$ in the system (also expressed in number of spins). Here we shall consider only large spin blocks $\B_L$, i.e. $L \gg \xi$, and study the dependence of entanglement, as given for instance by the saturated entropy $S(\gamma,\lambda)$, on the hamiltonian parameters $\gamma$ and $\lambda$.

Our characterization of entanglement loss along RG trajectories will progress through three 
stages, refining at every step the underlying
ordering of quantum correlations.
($i$) First, we review the observation that the vacuum corresponding to the ultraviolet fixed point of a theory is more entangled than the vacuum corresponding to its infrared fixed point \cite{VLRK02}, indicating a global loss of entanglement. ($ii$) Second, for the quantum Ising model we detect a monotonic decrease of the saturation entropy {\em along} the RG flows, that is, we see that part of the entanglement in the theory is lost every time a RG transformation is applied. ($iii$) In the third stage we identify a fine-grained characterization of this monotonic loss of entanglement by unveiling a rigid reordering of the eigenvalues of $\rho_L$ along RG trajectories. We show that the above decrease in entropy actually follows from a much more demanding set of inequalities for the eigenvalues of $\rho_L$, known as majorization relations, that are also fulfilled along the RG flow.

{\sl Global loss of entanglement.---}
A RG flow interpolates between the ultraviolet fixed point (UV) of a theory and the infrared (IR) one. To prove its irreversibility it is
enough to construct an observable quantity at the 
critical points, typically called $c$, such that $c_{UV}\ge c_{IR}$. Any
observer presented with two different states of a quantum system
could measure $c$ in both and decide which state is the UV precursor
and which is the IR result. 
Zamolodchikov \cite{Zam86} constructed an observable quantity $C(g^i,\mu)$ 
for one dimensional quantum systems depending on the couplings $g^i$
of the theory and a subtraction point $\mu$ 
such that it decreases along
RG flows,
\be
 -\beta^i{\partial\over \partial g^i} C\le 0 \ ,
\ee
where $\beta^i\equiv \mu{{\rm d}g^i /{\rm d}\mu}$
are the beta functions of the theory.
At critical points, conformal invariance is recovered 
and $C(g^{i*},\mu)=c$ where $c$ is the central charge of 
the conformal field theory describing  universal properties
of the critical system.

The computation of the von Neumann entropy for quantum spin chains presented in Ref.~\cite{VLRK02} (which recovered results from conformal field theory first found in Ref.~\cite{HLW94}) showed that, at a quantum critical point, the entropy of a spin block $\B_L$ does not reach a saturation value for large $L$, but that it instead scales as
\be
 S_L(\gamma^*,\lambda^*) \approx {c+\bar c\over 6} \log_2 L \ ,
\ee
where $\bar c=c$ for spin chains and $c=1/2$ for the quantum Ising model. This 
universal result combines
with the c-theorem to guarantee irreversibility of the RG flows
for spin chains since the entropy turns out to be proportional
to the central charge at critical points, which are the initial
and end points of the trajectory. To be precise, the system
is probed fixing any large $L$. Then,
\be S_L^{UV} \ge S_L^{IR}\qquad \forall L \gg 1 .
\ee
Irreversibility is therefore rooted in an intrinsic property of the
vacuum. Note that the original proof of the c-theorem is 
based on correlators of the energy momentum tensor, which couple
to any degree of freedom. The entropy 
allows for a recasting of this constraint
in terms of vacuum properties only.

\begin{figure}
\epsfig{file=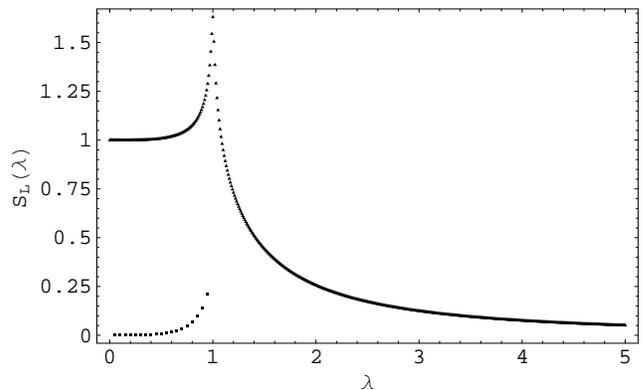,width=8.4cm}
\caption{Entropy $S_L(\lambda)$ of the quantum Ising chain as a function of the magnetic field  
$\lambda$ for $L = 100$. The lowest branch has $\epsilon > 0$.}
\label{fig:ising} 
\end{figure}

{\sl Monotonic loss of entanglement.---} Global entanglement loss
can be made point-wise along the RG flow. Here we illustrate this fact by considering the quantum ising chain, $\gamma=1$.
The computation of the saturated entropy $S(\lambda) = S(\gamma = 1,\lambda)$ is represented in Fig.~\ref{fig:ising}. 
The plot shows monotonic decrease of the entropy as $|1-\lambda|$ departs from zero, 
both for $\lambda > 1$ and $\lambda < 1$. In the first case the theory flows towards 
a product state corresponding to an infinitely massive
fermion with $S(\lambda\to\infty)=0$. The second case entails two possibilities.

($i$) The addition of a small magnetic field $\epsilon \sum_i \sigma_i^x$, $\epsilon \ll 1$, to the Ising hamiltonian induces again a flow towards a product state as $\lambda \rightarrow 0$, that is $S(\lambda\to 0)=0$. The effect of this magnetic field is to break the $Z_2$ symmetry of the model, thus emulating the spontaneous symmetry breaking that would occur in the corresponding field theory. The model with $\gamma=1, \lambda=1, \epsilon > 0$ was solved by Zamolodchikov showing that the spectrum develops eight masses \cite{Zam89}. 

($ii$) If symmetry breaking is not enforced, $\epsilon = 0$, then the flow leads to a Schr\"odinger's cat (or simply {\em cat}) state,
\be
\left(\ket{+}_1\ket{+}_2\cdots \ket{+}_N~~ + ~~
\ket{-}_1\ket{-}_2\cdots \ket{-}_N\right)/\sqrt{2},
\label{eq:cat}
\ee
where $\sigma_i^x\ket{\pm}_i = \pm\ket{\pm}_i$, so that $S(\lambda\to 0)=1$. 
While this state is a fixed point of the RG flow, it is
unstable with respect to $\epsilon$-deformations of the hamiltonian which 
makes it flow to a product state.  Since the {\em cat} state, whose entropy is 
$S(\rho_L)=1 \ \forall L<N$, does not obey scaling and violates the clustering
principle, only the spontaneously broken vacuum makes sense in field theory.

\begin{figure}
\epsfig{file=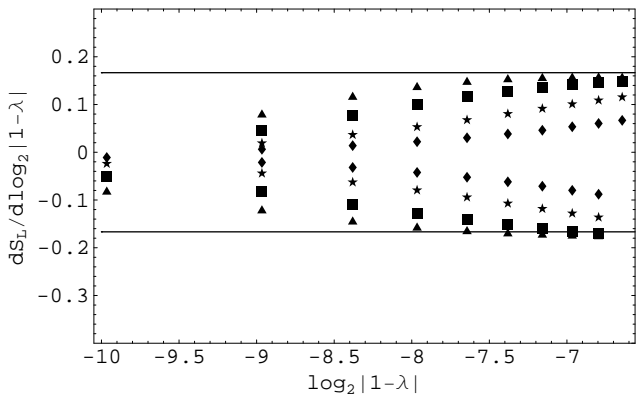,width=8.4cm}
\caption{Scaling of the Ising chain entropy $S_L(\lambda)$ as $\lambda$ approaches its
critical value $\lambda^* = 1$ ($\lambda\in(0.99,1.01)$), for various values of $L$ up to $L = 700$ (solid black triangles). 
Both left and right asymptotes scale as $S_L(\lambda)\sim -(1/6)\log_2 |1-\lambda|$.}
\label{fig:scaling} 
\end{figure}

The saturation entropy $S(\lambda)$ is particularly simple when the magnetic field $\lambda$ is close to the critical value $\lambda^* = 1$. Indeed, in this regime its scaling becomes symmetric around $\lambda = \lambda^*$:
\be 
S(\lambda)\sim -{1\over 6} \log_2|1-\lambda| ,
\ee
as can be seen in Fig.~\ref{fig:scaling}. Kitaev has previously derived this expression analytically \cite{Ki03}.

Let us  make clear that monotonicity is always 
present in one-coupling field theories.
 To see this, consider a theory with many coupling 
where the beta functions are expressed a
s gradient flows, that is
$ \beta^i=G^{ij}(g){\partial\over \partial_j}\Phi(g)$,
where $G^{ij}(g)$ is a positive matrix and $\Phi(g)$ is an observable. Then
 $ -\beta^i{\partial\over \partial g^i}\Phi(g)= 
-G^{ij}{\partial\Phi\over\partial g^i}{\partial\Phi\over\partial g^j}\le 0$,
proving monotonicity. This is a trivial statement for a 
one-coupling theory since the beta function 
only depends in one coupling and can always be written as the derivative of
some function.

{\sl Fine-grained loss of entanglement}.--- The monotonic decrease of the saturation entropy $S(\lambda)$ with growing value of $|1-\lambda|$ reflects the fact that the spectra of the underlying spin block density matrices become more and more ordered as we progress towards the infrared fixed point. It is natural to enquire if it is possible to characterize the reordering of density matrix eigenvalues along the RG flow beyond the simple entropic inequality discussed so far and thereby unveil some richer structure. 

Direct numerical inspection shows that the convergence of the entropy $S_L(\gamma,\lambda)$ in 
(\ref{eq:SL}) into a saturation value $S(\gamma,\lambda)$ for sufficiently large spin blocks $\B_L$, $L\gg \xi$, actually follows from the fact that the whole spectrum of $\rho_L(\gamma,\lambda)$ effectively converges into a limiting spectrum (up to contributions beyond the numerical accuracy $\epsilon_{acc} = 10^{-16}$ used in the calculations). Let $\vec{p}(\lambda) \equiv (p_1(\lambda), p_2(\lambda), \cdots, p_d(\lambda))$ denote the limiting spectrum for the Ising model with a given value $\lambda$ of the transverse magnetic field, where $p_1(\lambda)\geq p_2(\lambda) \cdots p_d(\lambda)\geq \epsilon_{acc}$. Then along both the RG trajectory associated to $\lambda>1$ and that associated to $\lambda<1$, we find the following result. If $|1-\lambda'| > |1-\lambda|$, then the spectrum $\vec{p}(\lambda')$ is more ordered than the spectrum $\vec{p}(\lambda)$ in the sense of the {\em majorization} relation, denoted  $\vec{p}(\lambda) \prec \vec{p}(\lambda')$ \cite{bhatia}. That is, we find that the set of $d$ inequalities
$\sum_{i=1}^{n}p_i(\lambda) \leq \sum_{i=1}^n p_i(\lambda'),   (n = 1,2,\dots,d)$
are simultaneously satisfied. This is a remarkable fact, since for large $d$, two randomly drawn probability distributions $\vec{x}$ and $\vec{y}$ are highly unlikely to fulfill all $d$ inequalities. 
Clearly, majorization reflects a very rigid reordering of spectra along the RG flow. The reordering given by majorization is actually so strict that it implies the monotonicity of a large number of popular entropies \cite{bhatia}. It follows, for instance, that all Renyi entropies of index $\alpha$ \cite{Renyi} also decrease monotonically along the RG-flow under study. This is not the first time majorization appears in the study of entanglement. In quantum information science, a whole theory of entanglement transformations under local operations and classical communication has been derived based on this relation (see \cite{NieVid} for a review).

Let us detail the above computations. As discussed in Ref.~\cite{VLRK02}, the density matrix $\rho_L(\lambda)$ results from the tensor product of the states $\varrho_j(\lambda)$ of $L$ fermion modes,
\be 
\rho_L(\lambda)=\varrho_1(\lambda)\otimes\dots\otimes\varrho_L(\lambda)\ ,
\ee
so that the spectrum of $\rho_L(\lambda)$ is the direct product of $L$ binary spectra,
\be
\left(\begin{array}{cc} q_{1}(\lambda) \\ 1-q_{1}(\lambda) \end{array} \right) 
\times
\left(\begin{array}{cc} q_{2}(\lambda) \\ 1-q_{2}(\lambda) \end{array} \right) 
\times \cdots \times
\left(\begin{array}{cc} q_{L}(\lambda) \\ 1-q_{L}(\lambda) \end{array}\right), 
\label{eq:spectrum}
\ee 
where we choose $q_{1}(\lambda) \geq q_{2}(\lambda) \geq \cdots \geq q_{L}(\lambda) \geq 1/2$. Then, for $\lambda > 1$ we find numerically that
\be
 \lambda < \lambda'~~ \Rightarrow ~~q_j (\lambda) \leq q_j(\lambda')   ~~~ j=1,\cdots, L,
\label{eq:mono}
\ee
where $L$ is larger than the correlation length $\xi$ for both $\lambda$ and $\lambda'$. That is, the spectrum of each individual fermion mode satisfies majorization, $\vec{q}_j(\lambda) \prec \vec{q}_j(\lambda')$. The majorization relation $\vec{p}(\lambda) \prec \vec{p}(\lambda')$ follows then from a recursive use of the following little lemma.

{\bf Lemma:} Let $\vec{x}$ and $\vec{y}$ denote two probability vectors and $\vec{z}$ a third one with components $z_{(ij)} \equiv x_iy_j$. Let the probability vectors $\vec{x}'$, $\vec{y}'$ and $\vec{z}'$ be related in the same way.
Then
\be
\left. \begin{array}{cc} 
\vec{x} \prec \vec{x}' \\ 
\vec{y} \prec \vec{y}' 
\end{array} \right\}
\Rightarrow \vec{z} \prec \vec{z}'.
\ee 

{\it Proof: The majorization relations $\vec{x} \prec \vec{x}'$ and $\vec{y} \prec \vec{y}'$ are equivalent \cite{bhatia} to the existence of doubly stochastic matrices $X$ and $Y$ such that $x_i = \sum_{k} X_{ik}x'_k$ and $y_j = \sum_{l} Y_{jl}y'_l$. It follows that 
\be
z_{(ij)} =  x_iy_j = \sum_{kl} X_{ik}Y_{jl} x'_k y'_l = \sum_{kl} Z_{(ij)(kl)}z'_{(kl)},
\ee
where $Z_{(ij)(kl)} \equiv X_{ik}Y_{jl}$ is also doubly stochastic and therefore $\vec{z} \prec \vec{z}'$.}

The case $\lambda < 1$ is slightly more complex. If a weak magnetic field is used to break the $Z_2$ symmetry of the Ising model, then the analytic derivation of Ref.~\cite{VLRK02} cannot be used. Instead, we have used DMRG techniques to directly compute the spectrum of $\rho_L(\lambda)$, obtaining that majorization is again satisfied. If, on the contrary, the $Z_2$ symmetry is preserved, then the spectrum of $\rho_L$ still decomposes as in (\ref{eq:spectrum}), but (\ref{eq:mono}) is violated by one fermionic mode and we cannot use the above lemma. Nevertheless, a careful numerical analysis for sufficiently large $L$ shows that majorization is once more fulfilled along the RG flow that converges to the cat state (\ref{eq:cat}).

\begin{figure}
\epsfig{file=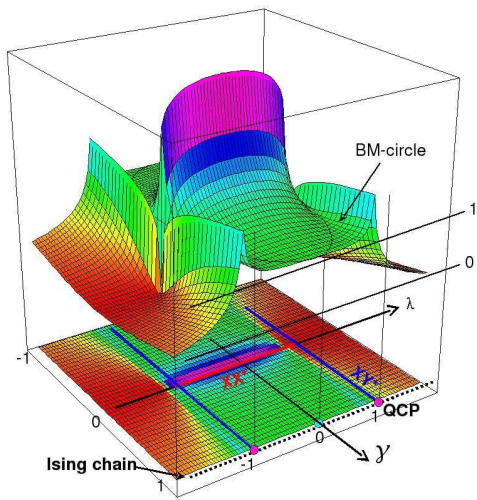,width=8.4cm}
\caption{Entropy $S_L(\gamma,\lambda)$ for the XY spin chain as a function of anisotropy $\gamma$ and magnetic field $\lambda$ for $L = 100$. The solid red line in the projection (phase diagram) is the critical XX-model (XX*), and the solid blue line is the critical XY-model (XY*). QCP is the quantum critical point of the Ising chain.}
\label{fig:xyentropy}
\end{figure}

The analysis of the rest of the XY model does not provide further 
insight into the continuum limit, since the region where $\gamma\not=1$
always corresponds to a massive fermion theory. (The coefficient
$\gamma$ is eaten by the normalization of the kinetic term.)
The result for the entropy $S_L(\gamma,\lambda)$ is 
represented in Fig.~\ref{fig:xyentropy}. It is interesting to note
that the saturated entropy takes the constant cat state value ($S_L = 1$) on the circle
$\gamma^2 + \lambda^2 = 1$ \cite{BmC71}. Furthermore, the entropy
changes by a subleading constant along the critical $\lambda$ line,
$S_L(\gamma,\lambda=1) - S_L(\gamma',\lambda=1)=(1/6) \log_2 (\gamma/\gamma')$. The jump from $c=1/2$ to $c=1$ that takes place at $\gamma=0$ 
is seen as a singularity in the field redefinition of the free fermion
to achieve the continuum limit. 

At any point in coupling space for the XY model that we have been able to check, the increase of entropy $S_{L'}(\gamma,\lambda)>S_{L}(\gamma,\lambda)$ for $L'>L$
is rooted in a strict reordering of the ground state:
\be
\vec{p}_{L'}(\gamma,\lambda) \prec \vec{p}_L(\gamma,\lambda).
\label{eq:L2}
\ee
This relation is valid only if $L$ is incremented in an even number of steps (i.e. at least
$L'=L+2$) due to the microscopic structure of the spin chain model. For large blocks of spins with 
$L \gg 1$, where adding two spins to a block can be regarded as an infinitesimal change in its size (for instance, in the continuum limit of a field theory), (\ref{eq:L2}) says that majorization also controls the variations in the spectrum of the vacuum density matrix when continuously increasing the block size.

In this letter we have presented a collection of numerical results
characterizing, in a simple quantum spin chain model, the fine-grained loss of entanglement along RG trajectories. This loss is a striking manifestation of the irreversibility of RG flows, one that only involves vacuum properties and not the whole hamiltonian of the system.

The idea to attack RG irreversibility from the state point of view rather than from a discussion of flows in the space of hamiltonians or correlators for the energy momentum tensor needs to be made more concrete and more general.
We have presented some evidence that RG irreversibility is already encoded in a delicate reordering of the ground state.  This work needs to be extended to other states and non-Gaussian theories.
If explicit and general requirements for physically acceptable RG transformations (as preserving unitarity) acting directly on quantum states can be worked out, then perhaps the monotonic character of entanglement loss can be derived from such considerations.

\end{document}